\documentclass[aps,pra,twocolumn,superscriptaddress,10pt]{revtex4-2}

\usepackage{graphicx}
\usepackage{amssymb}
\usepackage{bm}
\usepackage{booktabs}

\newcommand{\sgn}{\mathop{\mathrm{sgn}}}

\begin{document}

\title{A More Convex Ising Formulation of Max-3-Cut\\Using Higher-Order Spin Interactions}

\author{Robbe De Prins}
\email{robbe.deprins@ugent.be}
\affiliation{Photonics Research Group, Ghent University – imec, Technologiepark-Zwijnaarde 126, 9052 Gent, Belgium}
\affiliation{Applied Physics Research Group, Vrije Universiteit Brussel, Pleinlaan 2, 1050 Brussels, Belgium}

\author{Guy Van der Sande}
\affiliation{Applied Physics Research Group, Vrije Universiteit Brussel, Pleinlaan 2, 1050 Brussels, Belgium}

\author{Peter Bienstman}
\affiliation{Photonics Research Group, Ghent University – imec, Technologiepark-Zwijnaarde 126, 9052 Gent, Belgium}

\author{Thomas Van Vaerenbergh}
\affiliation{Large-Scale Integrated Photonics Lab, Hewlett Packard Labs, HPE Belgium, Diegem, Belgium}

\date{\today}

\begin{abstract}
Many combinatorial optimization problems (COPs) are naturally expressed using variables that take on more than two discrete values. To solve such problems using Ising machines (IMs)—specialized analog or digital devices designed to solve COPs efficiently—these multi-valued integers must be encoded using binary spin variables. A common approach is one-hot encoding, where each variable is represented by a group of spins constrained so that exactly one spin is in the “up” state. However, this encoding introduces energy barriers: changing an integer’s value requires flipping two spins and passing through an invalid intermediate state. This creates rugged energy landscapes that may hinder optimization. We propose a higher-order Ising formulation for Max-3-Cut, which is the smallest fundamental COP with multi-valued integer variables. Our formulation preserves valid configurations under single-spin updates. The resulting energy landscapes are smoother, and we show that this remains true even when the binary variables are relaxed to continuous values, making it well-suited for analog IMs as well. Benchmarking on such an IM, we find that the higher-order formulation leads to significantly faster solutions than the Ising baseline. Interestingly, we find that an empirical rescaling of some terms in the Ising formulation—a heuristic proposed in prior work—approaches the performance of the higher-order Ising formulation, underscoring the importance of empirical parameter tuning in COP encodings.
\end{abstract}

\maketitle

\section{\label{sec:intro}Introduction}
The challenge of solving a combinatorial optimization problem (COP) can be formulated as the task of finding the global minimum of an energy landscape. Many COPs are naturally expressed using multi-valued integer variables, i.e.~variables that can take one of $K$ discrete states. Examples can be found in statistical physics \cite{barahona1988application,de1995exact}, biology \cite{Protein_Folding, ComputationalBiology}, gas and power networks \cite{hojny2021mixed}, data clustering \cite{poland2006clustering}, scheduling \cite{Job_Scheduling,carlson1966scheduling}, and others \cite{Ising_formulations_of_many_NP_problems,dominguez2023encoding}.
Although hardware solvers that directly use multi-valued variables have been proposed, such as Potts machines \cite{potts_honari2020optical,potts_kalinin2018simulating}, most research focuses on binary-variable hardware, such as Ising machines (IMs)\cite{mohseni2022ising,inspiration_idea_thomas,inagaki2016coherent,pedretti2025solving,leleu2019destabilization}, adiabatic quantum computing \cite{farhi2000quantum}, and variational quantum approaches \cite{farhi2014quantum,hadfield2019quantum}.
This focus stems from the fact that any COP can be reformulated as an Ising problem or,  alternatively, as a Quadratic Unconstrained Binary Optimization (QUBO) problem, using only a polynomial overhead \cite{karp2009reducibility}.

To represent multi-valued integer variables within IMs, a common approach is one-hot encoding, where each $K$-state variable is represented by $K$ spins.
For each state $i \in \{1,\dots,K\}$, the corresponding spin configuration has the $i$-th spin set to 1 and all remaining spins are set to -1. 
In the case of a single isolated $K$-state variable, the Ising energy landscape is designed such that each of the $K$ one-hot configurations—corresponding to one valid state of the multi-valued integer—has the same minimal energy, while all other (invalid) configurations have higher energy.
Importantly, transitioning between valid states requires flipping two spins, and if these flips are performed sequentially, the system must pass through an invalid intermediate state. As a result, valid states are separated by energy barriers, creating a rugged landscape that complicates finding the global energy minimum and solving the COP.

In this work, we consider the Max-3-Cut problem, a variant of graph coloring and the smallest fundamental COP with multi-valued integer variables. We propose an alternative Max-3-Cut formulation that utilizes spin variables with higher-order interactions, i.e.~a higher-order Ising formulation (also known as Polynomial Uncostrained Spin Optimization, or PUSO). We will show that this formulation doubles the number of valid configurations while also allowing transitions between logical states via single-spin flips, thereby eliminating energy barriers. 

Our work is parallel to insights from recent studies on SAT problems, where transforming native higher-order formulations to quadratic alternatives was found to introduce a similar ruggedness \cite{dobrynin2024energy,hizzani2024memristor,pedretti2025solving,valiante2021computational}. Here, we reverse that path—transforming a quadratic formulation into a higher-order one—to recover smoother energy landscapes.

We further demonstrate that these structural benefits persist under continuous relaxations of the spin variables. This setting is relevant for analog IMs, which have attracted increasing attention in recent years \cite{inagaki2016coherent,bohm2021order,berloff2017realizing,king2018emulating,leleu2019destabilization,ercsey2011optimization,goto2019combinatorial,vadlamani2020physics,mohseni2022ising,reifenstein2023coherent}.

Finally, we benchmark both formulations using an analog IM inspired by simulated bifurcation dynamics \cite{inspiration_idea_thomas,higher_order_dSB}. The higher-order Ising formulation yields significantly faster solutions than the standard Ising formulation. 
Surprisingly, however, the latter can on average be brought within a factor of 2.75 of the higher-order formulation's speed through an empirical rescaling.
This rescaling approach was previously shown to be effective for a structure-based drug design problem \cite{original_meanAbsTrick_paper}, and in our earlier work, we showed that it is also effective for Max-3-Cut \cite{deprins2025ExternalFields}.
Here, we show that its effectiveness increases with problem size and that this cannot be fully explained by analyzing the individual components of the quadratic Ising formulation in isolation. We outline future directions to better understand and leverage this phenomenon.

\section{\label{sec:max3cut formulations}Max-3-Cut formulations}
The goal of the Max-3-Cut problem is to partition the vertices of an undirected graph into three disjoint sets while maximizing the number of edges connecting different sets. This can be viewed as a type of graph coloring, where each vertex is assigned one of three colors -- e.g. red, green, or blue -- with the goal of minimizing the number of edges between vertices of the same color.

\subsection{Ising formulation}
In Ref.~\cite{Ising_formulations_of_many_NP_problems}, a quadratic formulation in terms of bit variables, i.e.~a QUBO formulation, of the Max-3-Cut problem is proposed as follows. Denote the sets of vertices and edges in a given graph as $V$ and $E$, respectively.
For every vertex $v \in V$, a triplet of binary variables is introduced, $x_{v,i}$ where $i \in \{1,2,3\}$, using one-hot encoding: $x_{v,i}$ equals 1 if vertex $v$ has color $i$, and 0 otherwise. 
The QUBO energy function is defined as follows:
\begin{equation}
\mathcal{H}_{\text{QUBO}} = A \sum_{v \in V}\left(1-\sum_{i=1}^3 x_{v, i}\right)^2
+ B \sum_{(uv) \in E} \sum_{i=1}^3 x_{u, i} x_{v, i},
\label{eq:mapping lucas bits}
\end{equation}
where A and B are positive scalars. The first term enforces that all variable triplets are one-hot encoded, since this positive term only vanishes if every triplet contains a single 1 and two 0's, ensuring the vertex colors are well defined. The second term adds an energy penalty for every edge that connects vertices with the same color, aligning with the Max-3-Cut objective. The ratio $B/A$ determines the relative emphasis on valid color assignments versus cut maximization.

We convert the bit variables $x_{u,i} \in \{0,1\}$ to spin variables $\sigma_{u,i} \in \{-1,1\}$ as follows:
\begin{equation}
    x_i = \frac{\sigma_i+1}{2},
    \label{eq: transfo bits to spin}
\end{equation}
which yields the following Ising formulation:
\begin{eqnarray}
\mathcal{H}_{\text{Ising}} 
&=& \frac{A}{4} \sum_v \sum_{i \neq j} \sigma_{v, i} \sigma_{v, j} 
+ \frac{B}{4} \sum_{(uv)\in E} \sum_{i=1}^{3} \sigma_{u, i} \sigma_{v,i} \nonumber \\
&& + \sum_v \sum_{i=1}^{3} \left( \frac{A}{2} + \frac{B}{4} \text{deg}(v)\right) \sigma_{v, i},
\label{eq:mapping lucas spins}
\end{eqnarray}
where $\text{deg}(v)$ denotes the degree of vertex $v$.  In deriving Eq.~\ref{eq:mapping lucas spins} from Eq.~\ref{eq:mapping lucas bits}, we omitted constant terms that do not affect the energy landscape.

The allowed spin configurations, each representing a possible vertex color, are defined as follows:
\begin{equation}
\left\{
\begin{array}{ll}
\uparrow \downarrow \downarrow & \equiv~\text{red}, \\
\downarrow \uparrow \downarrow & \equiv~\text{green}, \\
\downarrow \downarrow \uparrow & \equiv~\text{blue}.
\end{array}
\right.
\label{eq:allowed states QUBO}
\end{equation}
Note that that the remaining (invalid) spin configurations occupy 62.5\% of the configuration space. Moreover, transitioning from one valid state to another requires two spin flips, thereby passing through an invalid intermediate state ($\downarrow\downarrow\downarrow$, $\downarrow\uparrow\uparrow$, $\uparrow\downarrow\uparrow$, or $\uparrow\uparrow\downarrow$) that forms an energy barrier.

\subsection{Higher-order Ising formulation}

We now propose an alternative Max-3-Cut formulation using higher-order spin interactions:
\begin{equation}
\mathcal{H}_{\text{HO}} = A \sum_v \sum_{i \neq j} \sigma_{v,i} \: \sigma_{v,j} 
+ B \sum_{(uv) \in E} \sum_{i \neq j} \sigma_{u,i} \: \sigma_{v,i} \: \sigma_{u, j} \: \sigma_{v, j}.
\label{eq: O24 mapping}
\end{equation}
Moreover, we increase the set of allowed configurations as follows:
\begin{equation}
\left\{
\begin{array}{l}
\uparrow \downarrow \downarrow \;\; \equiv\;\; \downarrow \uparrow \uparrow \;\; \equiv\; \text{red}, \\
\downarrow \uparrow \downarrow \;\; \equiv\;\; \uparrow \downarrow \uparrow \;\; \equiv\; \text{green}, \\
\downarrow \downarrow \uparrow \;\; \equiv\;\; \uparrow \uparrow \downarrow \;\; \equiv\; \text{blue}.
\end{array}
\right.
\label{eq:allowed states PUBO}
\end{equation}
That is, spin triplets now follow one-hot encoding, modulo global spin inversion within each triplet.

It is easy to check that the first term of Eq.~\ref{eq: O24 mapping} stabilizes these states. Indeed, when evaluated on a single spin triplet, this term has a value of $-2A$ for all of the allowed states, while the invalid states $\uparrow\uparrow\uparrow$ and $\downarrow\downarrow\downarrow$ lead to a value of $6A$.
The second term of Eq.~\ref{eq: O24 mapping} reflects the Max-3-Cut objective. Indeed, when considering two connected vertices, it evaluates to $6B$ if the vertices have the same color, while it equals $-2B$ if those vertices have a different color.

Note that Eq.~\ref{eq:allowed states PUBO} excludes only the $\uparrow\uparrow\uparrow$ and $\downarrow\downarrow\downarrow$ configurations, so invalid states occupy just 25\% of the configuration space. Moreover, transitions between any two colors can be achieved with a single spin flip, thereby avoiding the energy barriers introduced by one-hot encoding in the Ising mapping.

\subsection{Rescaled Ising formulation}
Alongside the Ising formulation of Eq.~\ref{eq:mapping lucas spins}, we also consider a simple variant obtained by scaling the terms that are linear in the spin variables by a factor of 0.6:
\begin{eqnarray}
\mathcal{H}_{\text{Ising, res}} 
&=& \frac{A}{4} \sum_v \sum_{i \neq j} \sigma_{v, i} \sigma_{v, j} 
+ \frac{B}{4} \sum_{(uv)\in E} \sum_{i=1}^{3} \sigma_{u, i} \sigma_{v,i} \nonumber \\
&& + 0.6 \sum_v \sum_{i=1}^{3} \left( \frac{A}{2} + \frac{B}{4} \text{deg}(v)\right) \sigma_{v, i},
\label{eq:mapping lucas spins rescaled 0.6}
\end{eqnarray}

This rescaling was initially proposed in Ref.~\cite{original_meanAbsTrick_paper}, which found it to be beneficial for a structure-based drug design problem of which the mapping contains one-hot encoding constraints, similar to Eq.~\ref{eq:mapping lucas spins}, along with additional problem-specific constraints. In our previous work \cite{deprins2025ExternalFields}, we showed that the same rescaling is also effective for Max-3-Cut. 
We found that this is because the original formulation of Eq.~\ref{eq:mapping lucas spins} only yields a correct Max-3-Cut solution (i.e.~a valid ground state) for increasingly smaller values of $B/A$ as the graph size grows.
Progressively decreasing $B/A$ is impractical due to resolution limits of the interaction parameters, but it turns out that an empirical rescaling factor of $0.6$ offers a simple and effective workaround.

\section{Energy landscapes for analog spins}
\label{sec:energy landscapes analog pins}
As discussed above, the higher-order Ising formulation offers clear advantages over the quadratic Ising formulation when working with binary spin variables $\sigma_i \in \{-1,1\}$. Unlike the quadratic formulation, which excludes more than half of the configuration space and introduces energy barriers between feasible states, the higher-order formulation permits twice the number of configurations and smoother transitions via single spin flips, resulting in a less rugged energy landscape.  
In this section, we extend this comparison to the analog case by replacing binary spins $\sigma_i$ with continuous variables $s_i \in \mathbb{R}$. This relaxation is relevant in many combinatorial optimization solvers \cite{inagaki2016coherent,bohm2021order,berloff2017realizing,king2018emulating,leleu2019destabilization,ercsey2011optimization,goto2019combinatorial,vadlamani2020physics,mohseni2022ising,reifenstein2023coherent}.

To illustrate differences in energy landscapes, we visualize small-scale instances of the higher-order Ising formulation of Eq.~\ref{eq: O24 mapping} and the Ising formulation of Eq.~\ref{eq:mapping lucas spins}.
As detailed in Appendix~\ref{appendix:rescaled_landscape}, the rescaled Ising formulation of Eq.~\ref{eq:mapping lucas spins rescaled 0.6} produces landscapes similar to the original Ising formulation for these instances. We therefore focus here on comparing the higher-order formulation to the standard quadratic formulation. Visualizations for the rescaled case are provided in Appendix~\ref{appendix:rescaled_landscape}.

Fig.~\ref{fig:singleNode_3dplots_landscape}(a,b) visualizes the quadratic and higher-order landscapes for a single analog spin triplet, representing a single, unconnected graph vertex. I.e., we consider only the terms with prefactor $A$ from Eqs.~\ref{eq:mapping lucas spins} and \ref{eq: O24 mapping}, and normalize the energy values for direct comparison between the formulations.  
Panels (c) and (d) highlight the states with energy below an arbitrary threshold of -0.9, corresponding to the allowed spin configurations defined in Eqs.~\ref{eq:allowed states QUBO} and \ref{eq:allowed states PUBO}, respectively. In the Ising landscape (panel c), the allowed configurations are separated by energy barriers, resulting in a rugged energy landscape. In contrast, the higher-order Ising landscape (panel d) features a barrier-free path connecting the allowed states. 
Similar to the case of discrete spins, the higher-order formulation allows analog spins to access a larger portion of the configuration space, where transitions between allowed states can occur without overcoming energy barriers.

\begin{figure}[htb]
    \includegraphics[width=1.0\linewidth]{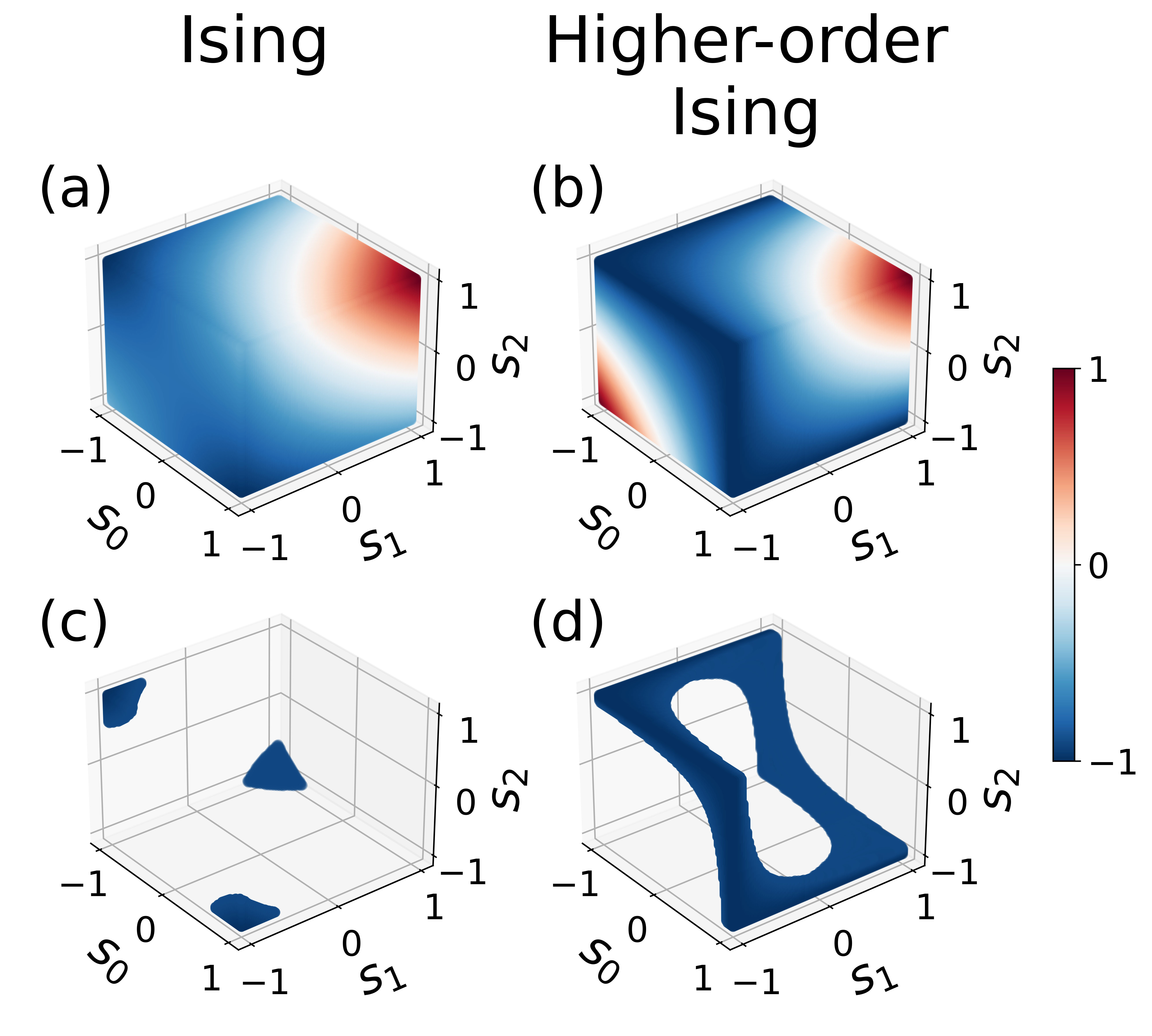}
    \caption{\label{fig:singleNode_3dplots_landscape}
     Normalized energy landscapes for a single vertex, represented by a spin triplet,  under the (a) Ising and (b) higher-order Ising formulations. Panels (c) and (d) highlight states with energy below –0.9 for these respective formulations. For the Ising formulation, the lowest-energy states  ($\uparrow\downarrow\downarrow$, $\downarrow\uparrow\downarrow$, $\downarrow\downarrow\uparrow$) are separated by energy barriers. In contrast, the minimal energy states for the higher-order Ising formulation  ($\uparrow\downarrow\downarrow$, $\downarrow\uparrow\downarrow$, $\downarrow\downarrow\uparrow$, $\downarrow\uparrow\uparrow$, $\uparrow\downarrow\uparrow$, $\uparrow\uparrow\downarrow$) are connected by a flat, barrier-free energy path.
    }
\end{figure}

\begin{figure*}
    \includegraphics[width=1.0\linewidth]{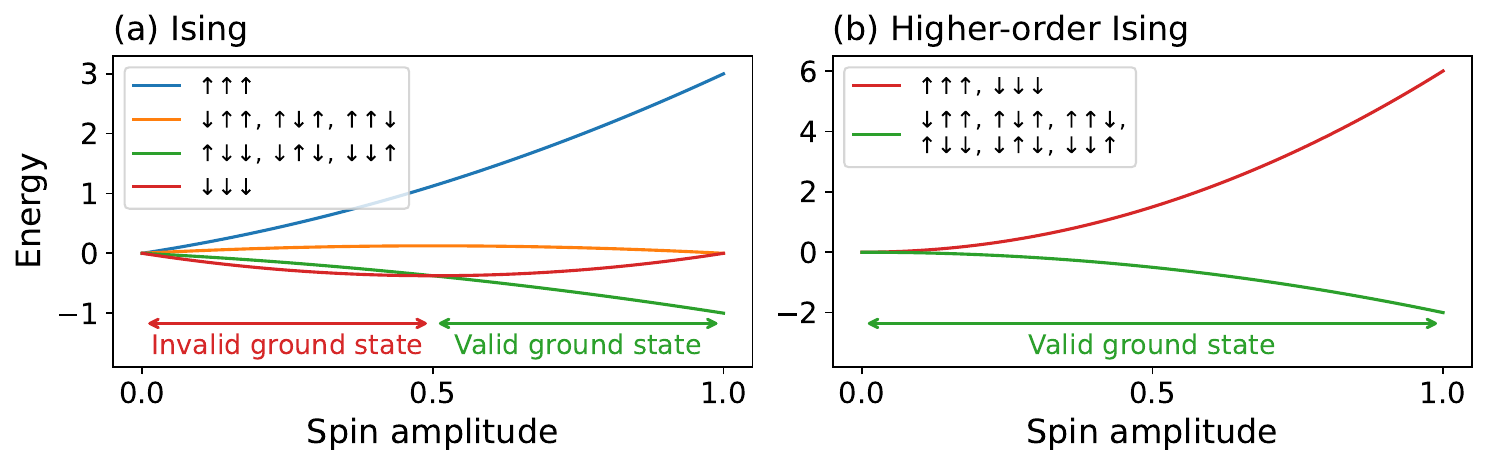}
    \caption{\label{fig:singleNode_2dplots_homogeneous}
    Energy landscape for a single vertex, represented by a spin triplet, assuming equal spin amplitudes. For the Ising formulation (a), the lowest energy configuration is only correctly one-hot encoded at sufficiently large spin amplitudes. For smaller amplitudes, the ground state ($\downarrow\downarrow\downarrow$) violates this constraint. 
    In contrast, the higher-order Ising formulation (b) yields a valid ground state for all amplitude values.
    }
\end{figure*}

It is well known that amplitude inhomogeneity can reduce the performance of analog solvers, such as analog IMs, since it may lead to an improper mapping of the COPs they aim to solve \cite{leleu2017combinatorial}. To counteract this, many of these solvers include mechanisms specifically aimed at suppressing such inhomogeneities \cite{leleu2019destabilization,bohm2021order}.
With this in mind, we consider paths for which the three spins have equal amplitudes, i.e.~$(\pm s,\pm s,\pm s), \: \forall s \in [0,1]$. In Fig.~\ref{fig:singleNode_3dplots_landscape}, these paths extend from the origin to the corners of the cube.
Fig.~\ref{fig:singleNode_2dplots_homogeneous}(a) shows the Ising energy of Eq.~\ref{eq:mapping lucas spins} along these paths. For large spin amplitudes ($s>0.5$), we observe the desired behaviour: the three valid one-hot encoded states of Eq.~\ref{eq:allowed states QUBO} are energetically degenerate, and they are lower in energy than the invalid configurations. At lower amplitudes ($s<0.5$), however, the $\downarrow\downarrow\downarrow$ configuration, which violates the one-hot encoding constraint, is the ground state.
This arises from the scaling mismatch between the linear and quadratic terms in Eq.~\ref{eq:mapping lucas spins}, which makes the linear terms dominate over the quadratic terms for small spin amplitudes. Since these linear terms are positive, the energy is minimized by pointing the spins down.
This presents a challenge because many analog solvers initialize spin amplitudes near zero \cite{yamamura2024geometric,inspiration_idea_thomas,inagaki2016coherent,leleu2019destabilization}. As a result, the system may initially favor the incorrect $\downarrow\downarrow\downarrow$ configuration, which can act as a distractor during solver operation. As shown in more detail in our recent work \cite{deprins2025ExternalFields}, such imbalances can indeed degrade performance if not properly mitigated in the solver's dynamics.
Fig.~\ref{fig:singleNode_2dplots_homogeneous}(b) shows the higher-order Ising energy of Eq.~\ref{eq: O24 mapping} along the paths of equal spin amplitudes.
In contrast to the Ising formulation, we observe that the valid states of Eq.~\ref{eq:allowed states PUBO} are energetically favourable for all values of the spin amplitude $s$. In other words, we observe no imbalances for the higher-order Ising formulation, which results from the fact that the suppression of invalid states ($\uparrow\uparrow\uparrow$ and $\downarrow\downarrow\downarrow$) is performed with quadratic terms only. 

\begin{figure*}
    \includegraphics[width=1.0\linewidth]{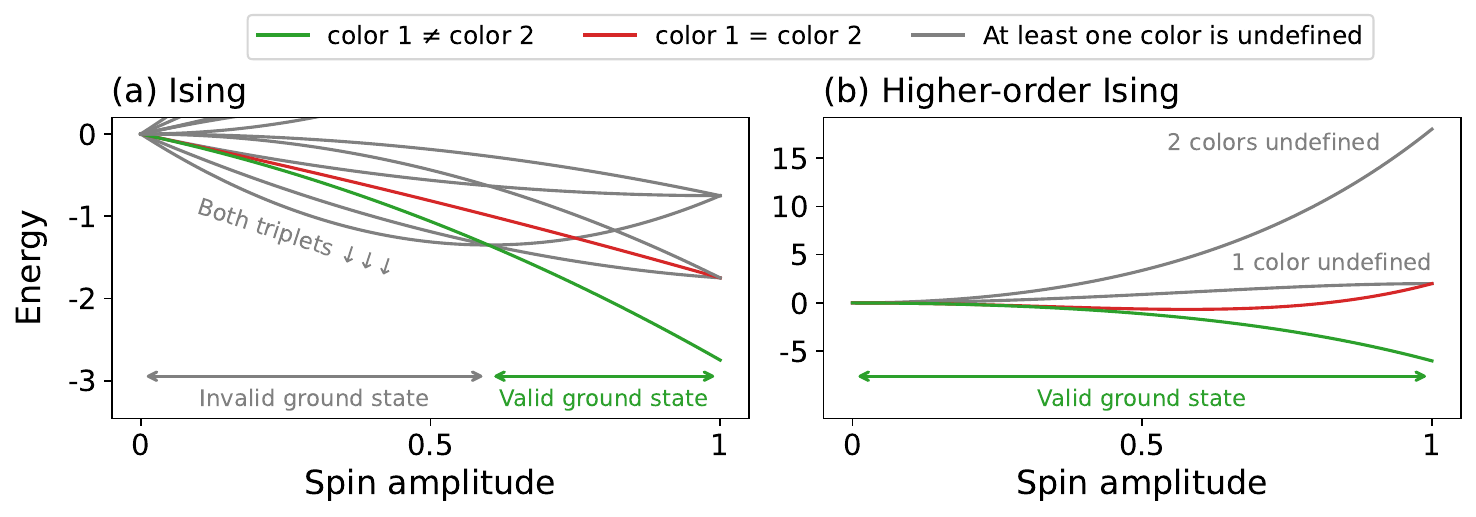}
    \caption{\label{fig:2connectedNodes_2dplots_homogeneous}
    Energy landscape for two connected vertices, assuming equal spin amplitudes. (a) For the Ising formulation, the lowest energy configuration corresponds to a correct Max-3-Cut solution (green) only at sufficiently large spin amplitudes. For smaller amplitudes, the ground state is $\downarrow\downarrow\downarrow$ for both vertices, violating the one-hot encoding constraint. (b) For the higher-order Ising formulation, the ground state solves the problem regardless of the value of the spin amplitude. In both cases, $B/A=1$ is used. In Appendix~\ref{appendix:vary_B_over_A}, we show that different values of $B/A$ yield similar energy landscapes.
    }
\end{figure*}

We now extend our discussion to the case of two connected vertices, each represented by a spin triplet. 
In Fig.~\ref{fig:2connectedNodes_2dplots_homogeneous}(a), we visualize the lowest-energy spin configurations of the Ising formulation,  under the assumption of homogeneous spin amplitudes, and for $B/A=1$.
The green curve corresponds to a correct one-hot encoding for both triplets, with the vertices assigned different colors, thus solving the Max-3-Cut problem.
However, as in Fig.~\ref{fig:singleNode_2dplots_homogeneous}(a), this configuration only becomes energetically favorable at sufficiently large spin amplitudes ($s>0.6$). For smaller amplitudes ($s<0.6$), it is favorable to set both triplets to the invalid $\downarrow\downarrow\downarrow$ state. 
As shown in Appendix~\ref{appendix:vary_B_over_A}, this behavior persists for other values of $B/A$, with the transition between ground states occurring at different values of $s$.
As discussed before, this preference for $\downarrow\downarrow\downarrow$ triplets at small amplitudes results from the linear terms outweighing the quadratic terms in the Ising formulation of Eq.~\ref{eq:mapping lucas spins} at small spin amplitudes. 

One might expect a similar issue in the higher-order Ising formulation of Eq.~\ref{eq: O24 mapping}, where the quadratic terms could dominate the fourth-order ones at small amplitudes. However, Fig.~\ref{fig:2connectedNodes_2dplots_homogeneous}(b) shows that the Max-3-Cut solution (green curve) remains the ground state across all values of $s$. In Appendix~\ref{appendix:vary_B_over_A}, we further show that this holds for any value of $B/A > 0$.
While this observation is limited to the case of two connected vertices, the underlying structure of Eq.~\ref{eq: O24 mapping} offers insight into its broader potential. The quadratic and fourth-order terms play different roles: the former enforces one-hot encoding (scaled by A), while the latter promotes the Max-3-Cut objective (scaled by B). 
Hence, any imbalance in the relative strength of the quadratic and fourth-order terms essentially modifies the effective value of $B/A$. Since $B/A$ is a hyperparameter, optimizing its value can partly compensate for these imbalances. 
In contrast, for the Ising formulation of Eq.~\ref{eq:mapping lucas spins}, both the one-hot encoding constraint and the Max-3-Cut objective are enforced using a combination of linear and quadratic terms. Because the linear terms are strictly positive, imbalances distort the energy landscape more drastically, making the $\downarrow\downarrow\downarrow$ configuration energetically favorable even though it violates the intended encoding.

\section{Performance comparison of the formulations}
\label{sec:results}
In this section, we compare the performance of the different Max-3-Cut formulations introduced in Section~\ref{sec:max3cut formulations} by applying them to a set of graphs generated using the rudy generator \cite{rudy_graph_generator}. For each graph size (5, 10, 20, 30, 40, 50, and 60 vertices), we consider 10 graph instances with an edge probability of 0.5 \cite{deprins2025rudygraphs,BiqMac}.

The formulations are benchmarked on an analog Ising machine, where the spin dynamics for each spin $s_i$ are governed by:
\begin{equation}
    \frac{ds_i}{dt} = -s_i + \tanh\left(\alpha s_i + \beta I_i\right),
    \label{eq: sigmoid nonlin}
\end{equation}
where $\alpha$ is the linear gain. $\beta$ is the interaction strength, which follows a commonly used linear annealing scheme \cite{paper_Ganguli,PaperJacob_UsingContinuationMethods} (see Appendix~\ref{appendix:methods} for more details). $I_i$ is the local field of $s_i$, which is further modeled as:
\begin{eqnarray}
I_i = 
\left\{
\begin{array}{ll}
J_i^{(1)} + \sum_j J_{ij}^{(2)} \, \sgn(s_j), 
& \text{for Ising,} \\[12pt]
\begin{array}{l}
\sum_j J_{ij}^{(2)} \, \sgn(s_j) \\
+ \sum_{j<k<l} J_{ijkl}^{(4)} \, \sgn(s_j s_k s_l),
\end{array}
& \shortstack[l]{for higher-\\order Ising.}
\end{array}
\right.
\label{eq:spin sign method cases}
\end{eqnarray}
Here, $\text{sgn}(\cdot)$ denotes the sign function. We adopt this local field model, which is inspired by the simulated bifurcation algorithm \cite{inspiration_idea_thomas,higher_order_dSB}, because it effectively incorporates interactions of different orders \cite{deprins2025ExternalFields,deprins2025HigherOrder}. As discussed in the previous section, mixed-order interactions can introduce imbalances that degrade solver performance. The use of the sign function in Eq.~\ref{eq:spin sign method cases} mitigates this issue by ensuring that higher-order and lower-order contributions are treated consistently, preventing any one interaction order from disproportionately dominating the dynamics. While, for the small-scale problem discussed in the previous section, we demonstrated that such imbalances are more pronounced for the Ising formulation than for higher-order formulation (cf.~Figs.~\ref{fig:singleNode_2dplots_homogeneous} and \ref{fig:2connectedNodes_2dplots_homogeneous}), they are also expected to arise in the higher-order case as problem size increases. Hence, we employ the spin sign method for all formulations in the following benchmark.

The IM is simulated by numerically integrating Eqs.~\ref{eq: sigmoid nonlin} and \ref{eq:spin sign method cases} via the Euler–Maruyama method, which includes stochastic noise (see Appendix~\ref{appendix:methods} for details).
We compare the formulations of Section~\ref{sec:max3cut formulations} in terms of time-to-solution (TTS):
\begin{eqnarray}
\text{TTS} = \left\{
\begin{array}{ll}
T, & \text{if } P > 0.99, \\[6pt]
T \frac{\log(0.01)}{\log(1 - P)}, & \text{if } 0 < P \leq 0.99, \\[6pt]
\infty, & \text{if } P = 0,
\end{array}
\right.
\label{eq:TTS}
\end{eqnarray}
which denotes the time needed to reach the target state with 99\% probability. Here $T$ is the (dimensionless) time window over which Eq.~\ref{eq: sigmoid nonlin} is integrated, and $P$ is the probability of reaching the optimal solution--as obtained using the publicly available \texttt{max\_k\_cut} solver \cite{MaxKcut_solver_Fakhimi}--within that time window. For each problem instance and for each set of hyperparameters (cf.~Appendix~\ref{appendix:methods}), $T \in [0,10^4]$ is selected to minimize the resulting TTS.

\begin{figure*}
    \includegraphics[width=1.0\linewidth]{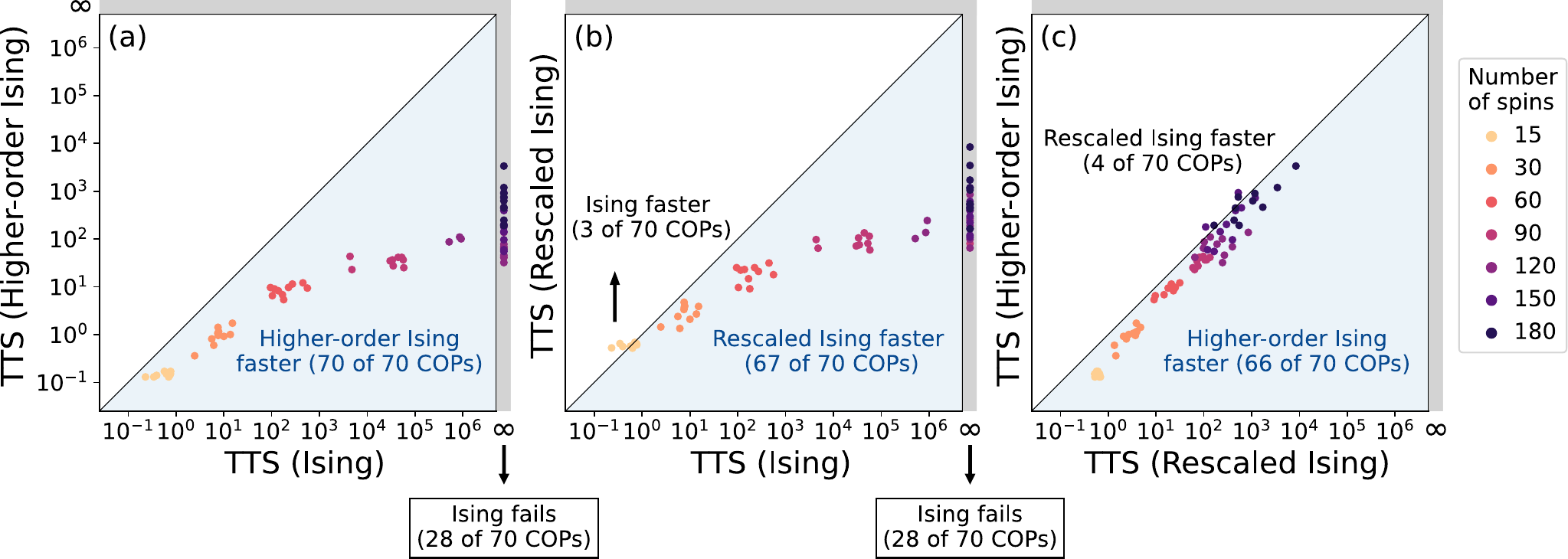}
    \caption{\label{fig:TTS comparison spinsign}Comparison of time-to-solution between the three formulations (Eqs.~\ref{eq:mapping lucas spins}, \ref{eq: O24 mapping}, and \ref{eq:mapping lucas spins rescaled 0.6}) of Max-3-Cut. Panels a and b show that both the higher-order Ising formulation and the rescaled Ising formulation significantly outperform the standard Ising formulation. The 28 dots in the shaded region on the right mark COPs that the standard Ising formulation failed to solve within the time limit $t_{\text{max}} = 10^4$. Panel c further indicates that the higher-order formulation generally leads to faster solutions than the rescaled quadratic one.}
\end{figure*}

In Fig.~\ref{fig:TTS comparison spinsign}, we compare the Max-3-Cut formulations of Section~\ref{sec:max3cut formulations} in terms of TTS. 
Fig.~\ref{fig:TTS comparison spinsign}(a) compares the Ising formulation of Eq.~\ref{eq:mapping lucas spins} with the higher-order formulation of Eq.~\ref{eq: O24 mapping}. All data points reside below the diagonal (light blue region), indicating that all problems are solved faster using the higher-order formulation than using the standard quadratic one. Moreover, the latter formulation failed to solve 28 out of 70 COP instances for any choice of hyperparameters (cf.~Appendix~\ref{appendix:methods}) within the maximum allowed time $t_{\text{max}}=10^4$, resulting in a success rate of zero and $\text{TTS}=\infty$. Since the higher-order formulation did succeed on these 28 instances, yielding finite TTS values, these points appear in the grey region on the right side of the figure. This highlights the clear advantage of the higher-order formulation over the original Ising formulation.

A similar picture is obtained in Fig.~\ref{fig:TTS comparison spinsign}(b), which compares the original Ising formulation with its rescaled variant from Eq.~\ref{eq:mapping lucas spins rescaled 0.6}. Here, only 3 instances favor the unscaled Ising formulation, while the rescaled version performs better on all others. As in panel (a), 28 COPs lie in the grey area on the right, underscoring the inferior performance of the original Ising formulation.

Finally, Fig.~\ref{fig:TTS comparison spinsign}(c) directly compares the higher-order and rescaled Ising formulations. The higher-order formulation solves 66 out of 70 instances faster, demonstrating a consistent performance advantage. On average, it is $2.75 \pm 1.39$ times faster than the rescaled Ising formulation.

The strong performance of the higher-order formulation compared to the original Ising formulation, as seen in panel (a), aligns with expectations: the higher-order formulation leverages a larger part of the configuration space and removes energy barriers between allowed vertex configurations. 
However, it is noteworthy that the empirical rescaling—achieved by simply multiplying the linear Ising terms by 0.6—makes the performance of the quadratic formulation approach that of the higher-order formulation, as shown in panel (c). In prior work \cite{deprins2025ExternalFields}, we showed that the rescaling with this value addresses the imbalance between the one-hot encoding constraints and the Max-3-Cut objective, ensuring that the prefactor ratio $B/A$ remains within a practical range. 
Still, one may expect the structural benefits of the higher-order formulation to yield a larger advantage over the rescaled Ising formulation, since all energy landscapes in Section~\ref{sec:energy landscapes analog pins} are similar for the original and rescaled Ising formulations (cf.~Appendix~\ref{appendix:rescaled_landscape}). 
This apparent contradiction is clarified by panel (b), where we see that the original and rescaled Ising variants work equally well for small problems (light-colored dots), and that the benefit of the factor of $0.6$ only appears for increasing problem size. Therefore, the effectiveness of rescaled Ising formulation cannot be fully understood by analyzing its building blocks in isolation, like we did in Section~\ref{sec:energy landscapes analog pins}.

\section{Discussion}
In this work, we introduced a higher-order Ising formulation for the Max-3-Cut problem. Unlike the commonly used Ising  formulation, which relies on one-hot encoding and yields rugged energy landscapes, the higher-order formulation enables smoother landscapes by allowing transitions between valid configurations through single-spin flips.
We showed that this structural advantage extends to the analog setting, where binary variables are relaxed to continuous spins, making the new formulation well-suited for analog solvers. 

Benchmarking on an analog IM confirmed that the higher-order formulation vastly outperforms the standard Ising formulation in terms of time-to-solution.
Further comparing to an empirically rescaled variant of the Ising formulation —a heuristic proposed in earlier work— showed that the higher-order formulation is on average $2.75 \pm 1.39$ times faster on the tested benchmark problems. This confirms that the improved convexity of the higher-order Ising landscape translates into meaningful performance gains.

However, it is surprising that the heuristic rescaling can narrow the performance gap with the structurally favourable higher-order formulation up to a constant factor, despite having little impact on the energy landscape of the small building blocks that compose the standard Ising formulation. It turns out that its beneficial effect only becomes apparent for larger graphs.

These findings illustrate a broader point: while mathematically sound mappings—such as those derived in Ref.~\cite{Ising_formulations_of_many_NP_problems}—are widely used, they are not necessarily optimal in practice. In the case of Max-3-Cut, the original Ising formulation works adequately for small instances, but requires rescaling to stay effective as the problem size increases. This points to the importance of empirical tuning.
However, relying on manual tuning also raises concerns about scalability and consistency.
Building on this, a natural next step is to move beyond manual tuning and toward more systematic data-driven strategies.
Recent and ongoing efforts have begun to explore how machine learning might uncover effective mappings from representative COP instances \cite{richoux2023automatic,richoux2023learning}. In this light, tailoring formulations to specific problem classes and solver architectures appears to be a promising next step.

Our findings suggest several concrete directions for future work. One direction is to evaluate how the performance of the higher-order and rescaled Ising formulations evolves with increasing problem size and varying graph density. Another is to gain a deeper understanding of the behavior of the rescaled Ising formulation on larger instances, using more advanced tools to visualize its energy landscapes \cite{dobrynin2024energy,masuda2025energy}. Such insights could inform further improvements to the higher-order formulation, potentially through a similar rescaling strategy.
Finally, it would be valuable to construct alternative formulations using different encoding strategies, such as binary encodings and domain-wall encodings \cite{dominguez2023encoding}.

Overall, our study underscores the interplay between theoretical formulation, energy landscape structure, and empirical solver performance—especially in analog settings—and opens up a broader design space for future combinatorial optimization methods.

\appendix
\section{\label{appendix:methods}Simulation of the analog Ising machine}
As detailed in Section~\ref{sec:results}, the temporal evolution of the analog IM is governed by Eqs.~\ref{eq: sigmoid nonlin} and \ref{eq:spin sign method cases}. To obtain the results in Fig.~\ref{fig:TTS comparison spinsign}, these equations are integrated via the Euler-Maruyama method:
\begin{equation}
\mathbf{s}_{t+1} = \mathbf{s}_t + \Delta t \left(-\mathbf{s}_t + \tanh\left(\alpha \mathbf{s}_t + \beta_t \mathbf{I}_t\right)\right) + \gamma \bm{\xi}_t,
\label{eq:euler_update}
\end{equation}
where $\mathbf{s}_t$ denotes the vector of spin amplitudes at time $t$, $\Delta t = 0.01$ is the time step, $\alpha$ is the linear gain, $\gamma=0.001$ is the noise strength and $\bm{\xi}_t$ is a vector of real values that are randomly drawn from a Gaussian distribution with zero mean and standard deviation of $\sqrt{\Delta t}=0.1$. $\mathbf{I}_t$ is the vector of local fields, as defined in Eq.~\ref{eq:spin sign method cases}. $\beta_t$ is the interaction strength, which follows a commonly used linear annealing scheme \cite{paper_Ganguli, PaperJacob_UsingContinuationMethods}:
\begin{equation}
    \beta_{t+1} = \beta_t + v_\beta\Delta t,
    \label{eq:linear_annealing}
\end{equation}
where $v_\beta$ is the annealing speed, and $\beta_0=0$.
The iterative updates of Eqs.~\ref{eq:euler_update} and \ref{eq:linear_annealing} proceed until the IM obtains the ground-state energy or, alternatively, until it completes $10^4/\Delta t$ steps (whichever comes first).

For each formulation (defined in Section~\ref{sec:max3cut formulations}) and for each problem instance, we conduct a grid search over the hyperparameters summarized in Table~\ref{tab:hyperparams}. For each hyperparameter configuration, the IM evolution is repeated 100 times to estimate the TTS, as defined in Eq.~\ref{eq:TTS}. The TTS values reported in Fig.~\ref{fig:TTS comparison spinsign} correspond to the configurations yielding the lowest TTS.

\begin{table*}
\caption{\label{tab:hyperparams}Hyperparameter grid used in the performance comparison of Section~\ref{sec:results}.}
\begin{ruledtabular}
\begin{tabular}{lccccc}
\textbf{Parameter} &
\shortstack{\textbf{Lowest} \\ \textbf{value}} &
\shortstack{\textbf{Highest} \\ \textbf{value}} &
\shortstack{\textbf{Spacing} \\ \textbf{type}} &
\shortstack{\textbf{Number} \\ \textbf{ of values}} \\
\midrule
Linear gain $\alpha$ & $-10$ & $1$ & Linear & 5 \\
Annealing speed $v_\beta$ & $10^{-5}$ & $10^{-1}$ & Logarithmic & 5 \\
Mapping parameter ratio $B/A$ (Ising) & $0$ & $180/N$ & Linear & 7 \\
Mapping parameter ratio $B/A$ (Higher-order Ising) & $10.5/N$ & $39/N - 0.1$ & Linear & 7 \\
\end{tabular}
\end{ruledtabular}
\end{table*}

\section{\label{appendix:rescaled_landscape}Analog energy landscapes for the rescaled Ising formulation}

The energy landscape visualizations for small graphs in Section~\ref{sec:energy landscapes analog pins} include both the standard Ising formulation from Eq.~\ref{eq:mapping lucas spins} and the higher-order Ising formulation of Eq.~\ref{eq: O24 mapping}. Here, we focus on the rescaled Ising formulation defined in Eq.~\ref{eq:mapping lucas spins rescaled 0.6} and demonstrate that it yields energy landscapes similar to those of the standard Ising mapping for these graphs. This observation is consistent with the performance trends shown in Fig.~\ref{fig:TTS comparison spinsign}(b), where the empirical rescaling only begins to affect performance as the problem size increases further.

\begin{figure*}
    \includegraphics[width=0.6\linewidth]{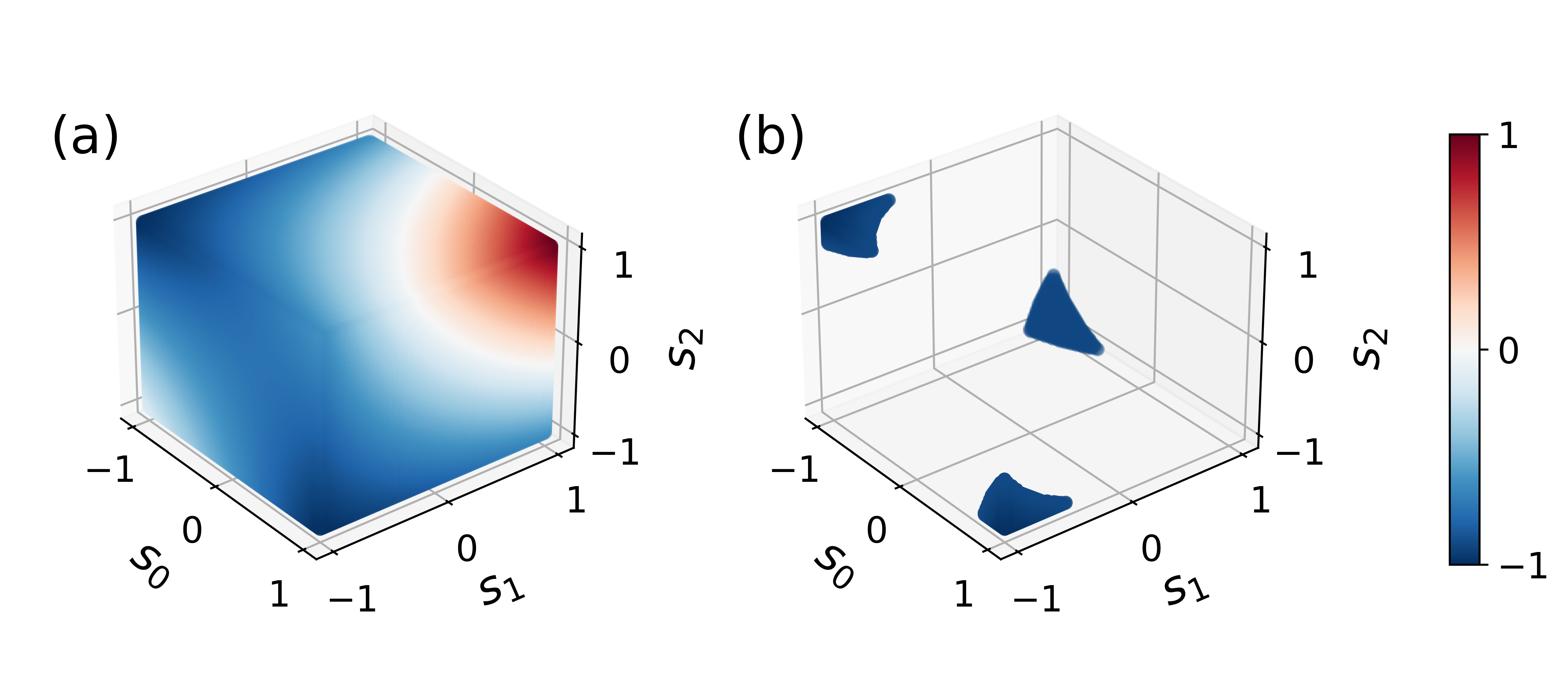}
    \caption{\label{fig:singleNode_3dplots_landscape_rescaledQUBO}
    Normalized energy landscapes for a single vertex, represented by a spin triplet, under the rescaled Ising formulation of Eq.~\ref{eq:mapping lucas spins rescaled 0.6}. Panel (b) highlights states with energy below –0.9. These figures look similar to Fig.~\ref{fig:singleNode_3dplots_landscape}(a,c), indicating that the rescaled Ising formulation and the standard Ising formulation behave similarly for this small problem.}
\end{figure*}

\begin{figure*}
    \includegraphics[width=1.0\linewidth]{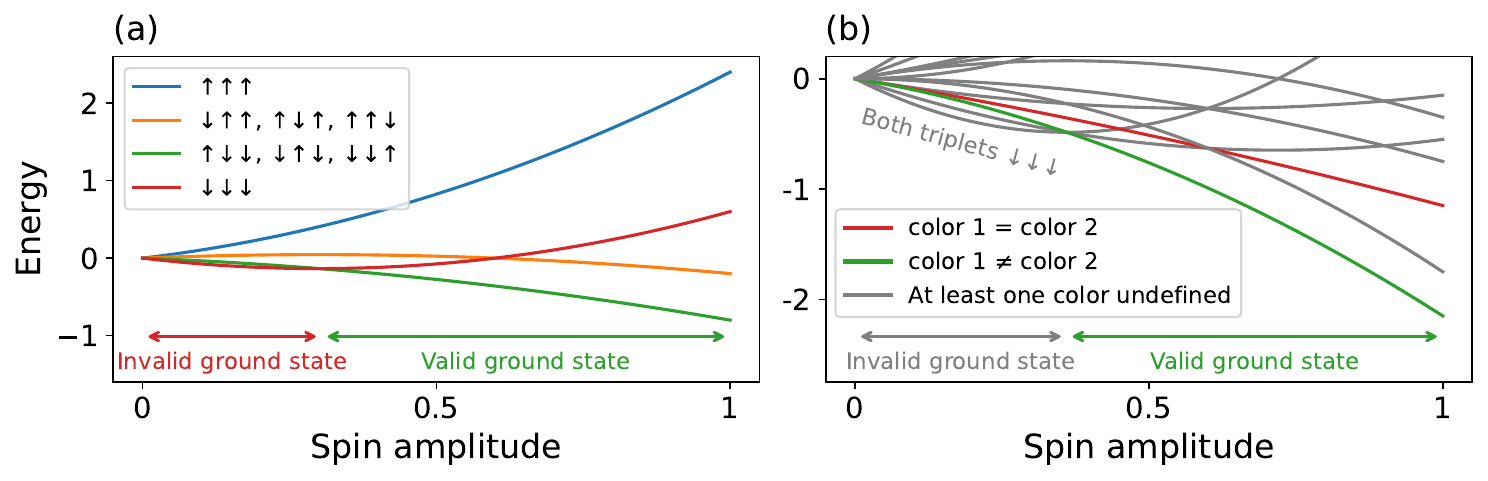}
    \caption{\label{fig:landscapes_homogeneousAmps_rescaledQUBO}
    Energy landscapes of the rescaled Ising formulation for (a) a single vertex, and (b) two connected vertices. We assume homogeneous spin amplitudes. For panel (b), $B/A=1$ is used. These figures look similar to Fig.~\ref{fig:singleNode_2dplots_homogeneous}(a) and Fig.~\ref{fig:2connectedNodes_2dplots_homogeneous}(a), indicating that the rescaled Ising formulation and the standard Ising formulation behave similarly for this small problem.}
\end{figure*}

Fig.~\ref{fig:singleNode_3dplots_landscape_rescaledQUBO} shows the energy landscape of a single spin triplet under the rescaled Ising formulation. The result closely resembles Fig.~\ref{fig:singleNode_3dplots_landscape}(a,c), indicating minimal qualitative differences between the two Ising variants at this scale. 
Also for the rescaled Ising formulation, Fig.~\ref{fig:landscapes_homogeneousAmps_rescaledQUBO} shows the energy landscapes for a single vertex (panel a) and two connected vertices (panel b) under the assumption of homogeneous spin amplitudes. Comparing to Fig.~\ref{fig:singleNode_2dplots_homogeneous}(a) and Fig.~\ref{fig:2connectedNodes_2dplots_homogeneous}(a) shows that the energy landscapes of both quadratic Ising variants are similar. The differences between the standard Ising formulation and the rescaled Ising variant can be understood as follows. First note that the linear terms in Eq.~\ref{eq:mapping lucas spins} are all strictly positive. This means these terms make it more energetically favourable for the spins to be in a down state. Going from Eq.~\ref{eq:mapping lucas spins} to Eq.~\ref{eq:mapping lucas spins rescaled 0.6}, the empirical rescaling weakens these linear terms by a factor of 0.6. Hence, down states are somewhat destabilized, and we observe in Fig.~\ref{fig:landscapes_homogeneousAmps_rescaledQUBO}(a) that the $\downarrow\downarrow\downarrow$ configuration (red curve) is destabilized with respect to its position in Fig.~\ref{fig:singleNode_2dplots_homogeneous}(a). Similarly, in Fig.~\ref{fig:landscapes_homogeneousAmps_rescaledQUBO}(b) we see that the curve with two $\downarrow\downarrow\downarrow$ configurations is destabilized with respect to Fig.~\ref{fig:2connectedNodes_2dplots_homogeneous}(a).

\section{\label{appendix:vary_B_over_A}Energy landscape for two connected nodes while varying $B/A$}
\begin{figure*}
    \includegraphics[width=1.0\linewidth]{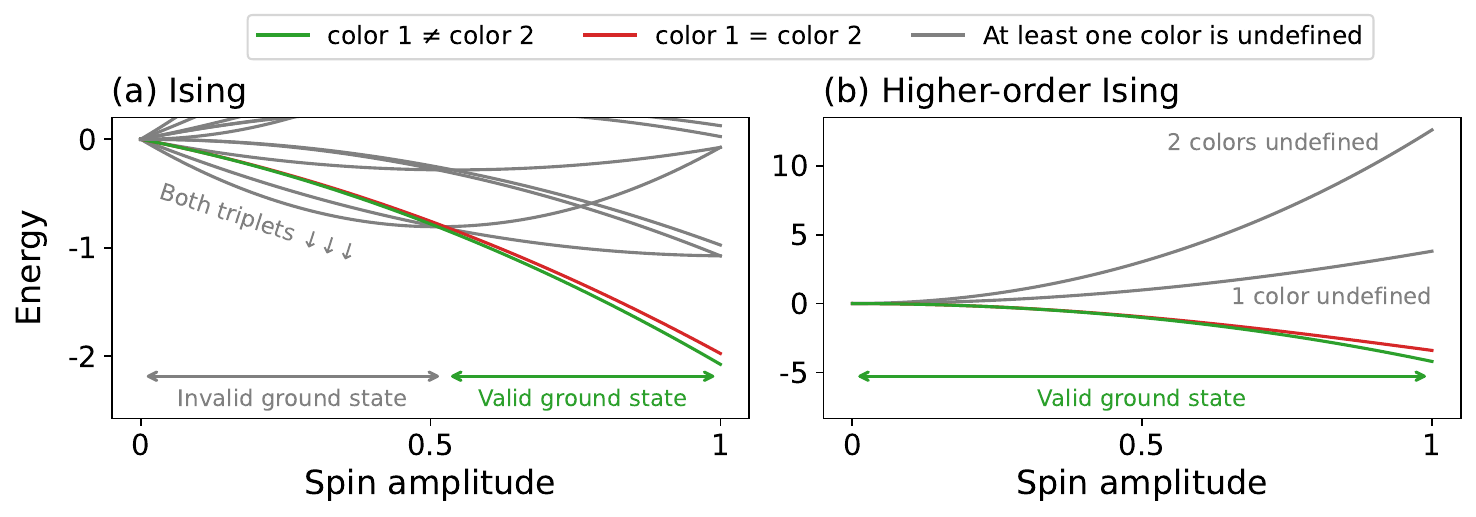}
    \caption{\label{fig:supply:B=0.1}
    Energy landscape for two connected vertices, assuming equal spin amplitudes. In contrast to Fig.~\ref{fig:2connectedNodes_2dplots_homogeneous}, this figure uses a value of $B/A=0.1$ for both formulations.
    }
    
\end{figure*}

\begin{figure*}
    \includegraphics[width=1.0\linewidth]{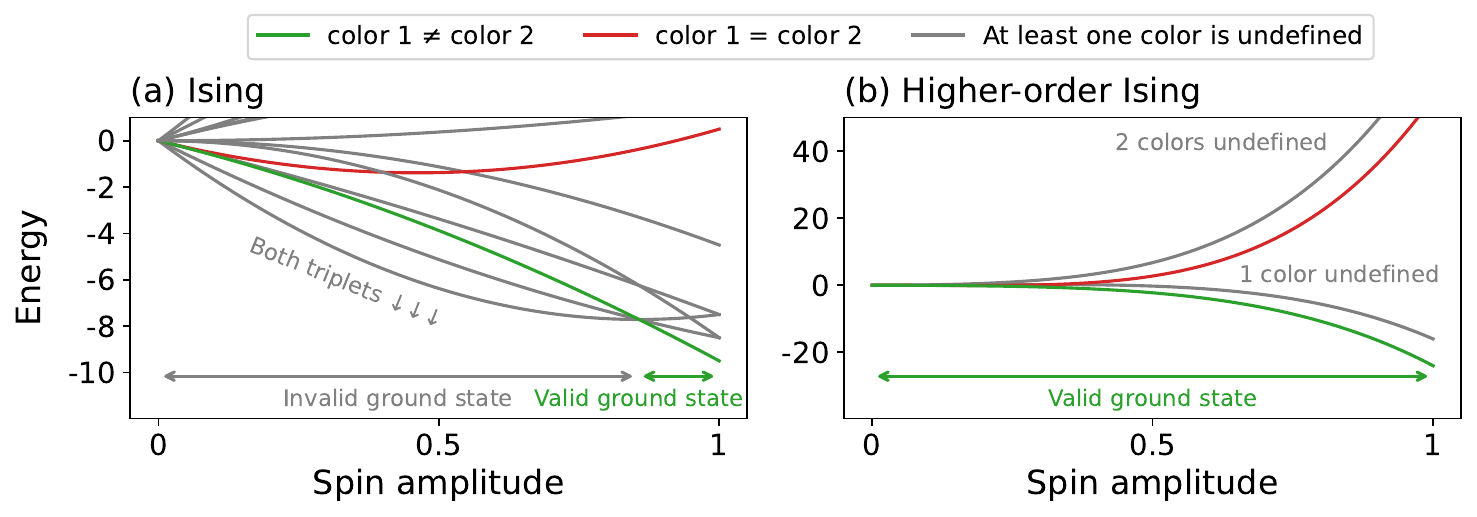}
    \caption{\label{fig:supply:B=10}
    Energy landscape for two connected vertices, assuming equal spin amplitudes. In contrast to Fig.~\ref{fig:2connectedNodes_2dplots_homogeneous}, this figure uses a value of $B/A=10$ for both formulations.
    }
\end{figure*}

Fig.~\ref{fig:2connectedNodes_2dplots_homogeneous} from the main text shows the energy landscape under the Ising formulation of Eq.~\ref{eq:mapping lucas spins} and the higher-order Ising formulation of Eq.~\ref{eq: O24 mapping} while assuming equal spin amplitudes. Moreover, this figure employs a value of $B/A=1$. Here we demonstrate how this figure changes when we vary $B/A$. Figs.~\ref{fig:supply:B=0.1} and \ref{fig:supply:B=10} visualize the landscape for $B/A=0.1$ and $B/A=10$, respectively. 

For the Ising formulation (Fig.~\ref{fig:supply:B=0.1}(a) and Fig.~\ref{fig:supply:B=10}(a)), we observe that under the new values of $B/A$, the same ground states remain present: for small spin amplitudes, the ground state is $\downarrow\downarrow\downarrow$ for both triplets, while for larger amplitudes, the Max-3-Cut solution (green line) is the ground state. However, the value of the spin amplitude where the ground state switches depends on $B/A$: larger values of $B/A$ shift the switching point to the right, leading to a larger region of incorrect ground state.

For the higher-order Ising formulation (Fig.~\ref{fig:supply:B=0.1}(b) and Fig.~\ref{fig:supply:B=10}(b)), the ground state corresponds to the Max-3-Cut solution, independent of the value of the spin amplitude. We now prove that this holds for any value of $B/A>0$ under the assumption of homogeneous amplitudes. By substituting all possible spin configurations $(\pm s,\pm s,\pm s), \forall s \in [0,1]$ in Eq.~\ref{eq: O24 mapping}, we end up with the following energies:
\begin{eqnarray}
\mathcal{H} = \left\{
\begin{array}{ll}
-4A s^2 - 2B s^4, & \text{if color 1 $\neq$ color 2}, \\[3pt]
-4A s^2 + 6B s^4, & \text{if color 1 $=$ color 2}, \\[3pt]
\phantom{-}4A s^2 - 2B s^4, & \text{if one color is undefined}, \\[3pt]
\phantom{-}12A s^2 + 6B s^4, & \text{if both colors are undefined}.
\end{array}
\right.
\label{eq:supply:cases energy 2 vertices PUBO}
\end{eqnarray}
It is easy to see that the first case in Eq.~\ref{eq:supply:cases energy 2 vertices PUBO}, corresponding to the Max-3-Cut solution where the two vertices have different colors, yields the lowest energy for all $s \in [0,1]$ and all $B/A>0$.

\section*{Data availability}
The authors declare that all relevant data are included in the manuscript. Additional data are available from the corresponding author upon reasonable request.\\

\section*{Author contributions}
R.D.P. performed the simulations and wrote the manuscript. G.V.d.S., P.B., and T.V.V. supervised the project. All authors discussed the results and reviewed the manuscript.\\

\begin{acknowledgments}
This research was funded by the Prometheus Horizon Europe project 101070195. It was also funded by the Research Foundation Flanders (FWO) under grants G028618N, G029519N, G0A6L25N, and G006020N. Additional funding was provided by the EOS project `Photonic Ising Machines'. This project (EOS number 40007536) has received funding from the FWO and F.R.S.-FNRS under the Excellence of Science (EOS) programme. The work was also partly supported by the Defense Advanced Research Projects Agency (DARPA) under Air Force Research Laboratory (AFRL) contract no FA8650-23-3-7313.
\end{acknowledgments}

\bibliography{references}

\end{document}